\documentclass[12pt]{article}
\usepackage{dcolumn}
\usepackage{bm}
\usepackage{graphics}
\usepackage{amsmath}
\usepackage{amssymb}
\usepackage{amscd}
\usepackage{afterpage}
\usepackage{float,times}
\usepackage{subfigure}
\usepackage{rotating}
\usepackage{multirow}
\usepackage{epsfig}
\usepackage{theorem}
\usepackage{moreverb}
\usepackage{euscript}
\usepackage{psfrag}

\begin{document}

\begin{titlepage}
\begin{center}

\bigskip

\vspace{6\baselineskip}

{\Large \bf

Alternative implementation of the Higgs boson \\}

\bigskip

\bigskip

\bigskip

{\bf Robert Foot and Archil Kobakhidze \\}

\smallskip

{ \small \it
School of Physics, The University of Melbourne, Victoria 3010, Australia
\\
E-mail: rfoot@unimelb.edu.au, archilk@unimelb.edu.au 
\\}

\vspace*{1.0cm}

{\bf Abstract}\\
\end{center}
\noindent
{\small 
We discuss an alternative implementation of the Higgs boson within the
Standard Model which is possible if the renormalizability condition is
relaxed. Namely, at energy scale $\Lambda$ the Higgs boson interacts at
tree-level only with matter fermions, while the full gauge invariance
is still maintained. The interactions with the electroweak gauge bosons
are induced at low energies through the radiative corrections. In this
scenario the Higgs boson can be arbitrarily heavy, interacting with the
Standard Model fields arbitrarily weakly. No violation of unitarity in
the scattering of longitudinal electroweak bosons occurs, since they
become unphysical degrees of freedom at energies $\Lambda \sim
\mathrm{TeV}$.  }

\bigskip

\bigskip

\end{titlepage}
\baselineskip=16pt


The Higgs particle \cite{Higgs:1964ia} 
is widely expected to be found in high energy experiments at the LHC. A theoretical motivation for the 
existence of such a particle is related to a ``bad''  high energy behaviour
of the electroweak theory with purely massive vector bosons. Namely,
scattering amplitudes of longitudinal modes ($W_L$ and $Z_L$) of the
massive electroweak vector bosons scale (at tree level)  as $\sqrt{s}$
with the center-of-mass energy ($E_{\rm CoM}=\sqrt{s}$), and thus
quickly saturate the unitarity bound at about $\sqrt{s}\sim 800$ GeV. A
relatively light (with a mass less than the unitarity bound) Higgs
boson, which couples to the electroweak gauge bosons in \emph{gauge
invariant} way, automatically provides cancellation of the bad energy
behaviour of the tree-level (and also higher order) amplitudes,
regaining the unitarity of the theory. The converse statement is also
true: any unitary and \emph{renormalizable} theory of massive
electroweak bosons must involve gauge invariant interactions with the
Higgs particle \cite{Cornwall:1974km}. If so, than the coupling of the
Higgs particle to electroweak vector bosons is \emph{fixed} by the gauge
invariance. This fact is crucial in searching for the Higgs particle in
actual experiments, e.g. in associate production of the Higgs and vector
bosons. 

In this paper we would like to explore an unorthodox model of spontaneous
electroweak symmetry breaking with nonstandard interactions of the Higgs
particle and electroweak vector bosons, which, to the best of our
knowledge, has not been discussed previously. The model we are going to
present is based on the observation that, contrary to a widespread
belief, gauge invariance \emph{per se} does not uniquely fix the
interactions of the Higgs boson. The standard interactions of the Higgs
boson with electroweak vector fields implied in the Standard Model,
follow only if, in addition to the gauge invariance, one demands the
\emph{renormalizability} of the theory. However, from a modern
perspective, the Standard Model is widely viewed as an effective
low-energy theory, and thus its renormalizability perhaps is not a very 
well justified requirement. If one relaxes the renormalizability
requirement, one can write down an alternative to the Standard Model
theory which is also perfectly unitary and gauge invariant, but the
interactions of the Higgs boson are entirely different. The key idea
behind such a theory is the following. If the coupling of the Higgs boson
to the electroweak gauge bosons is absent, then the
longitudinal modes are non-propagating degrees of  freedom due to the
gauge invariance in the classical theory at a certain energy scale 
$\Lambda$
\footnote{In \cite{Foot:2008tz} we have
discussed the model with the explicit breaking of the electroweak
symmetry where all the components of the electroweak doublet are not
propagating degrees of freedom at the classical level.}. 
The interactions with the electroweak gauge bosons and their masses 
are induced at low energies through the radiative corrections involving
fermionic loops.  
No violation of unitarity in
the scattering of longitudinal electroweak bosons occurs, since they
are unphysical degrees of freedom at energies $\Lambda \sim
\mathrm{TeV}$. The remarkable thing then is
that the unitarity bound is decoupled from the Higgs boson mass. That is
to say, the Higgs boson can be arbitrarily heavy, providing it
interacts
arbitrarily weakly with the Standard Model fields. In this
regime, the Higgs boson might not be observable at LHC.

We start by considering the electroweak doublet field $\Phi(x)$ in the
``polar'' paramaterization, 
\begin{equation}
\Phi(x)=\frac{1}{\sqrt{2}}H(x)\mathcal{X}(x)~,~~\mathcal{X}(x)= {\rm
e}^{-i\xi^a(x)\tau^a+i\xi^3(x)\mathbf{I}}
\left(
\begin{array}{c}
0 \\ 
1
\end{array} 
\right)\equiv U(\xi)\left(
\begin{array}{c}
0 \\ 
1
\end{array} \right)~,
\label{1}
\end{equation}
where $\tau^a$ $(a=1,2,3)$ are the half-Pauli matrices, and
$\mathbf{I}=\mathrm{diag}[1/2,1/2]$.  
The field $H(x)=(2\Phi^{\dagger}\Phi)^{1/2}$ is a modulus of the doublet
field (\ref{1})\footnote{The field $\mathcal{X}$ is unimodular,
$\mathcal{X}^{\dagger}\mathcal{X}=1$.} and thus it is an $SU(2)\times
U(1)_Y$-invariant component of $\Phi(x)$. The physical Higgs boson is
associated with quanta of $h(x)$, where $h(x)$ is a fluctuation over the
background vacuum expectation value $\langle H\rangle =v$, i.e.,
$H(x)=v+h(x)$.

Since the field $H(x)$ is $SU(2)\times U(1)_Y$-invariant, we can write a
gauge invariant Lagrangian solely for the $H(x)$ component of the
electroweak doublet field (\ref{1}) without invoking $SU(2)\times
U(1)_Y$ covariant derivatives as follows\footnote{$SU(2)\times
U(1)_Y$-invariant kinetic term for $H(x)$ can be written also in terms
of $\Phi(x)$: $$\frac{1}{2}\partial_{\mu}H\partial^{\mu}H\equiv
\partial_{\mu}\Phi^{\dagger}\partial^{\mu}\Phi+\frac{J_{\Phi~\mu}J_{\Phi}^{\mu}}{4(\Phi^{\dagger}\Phi)}~,~~\text{where}~,~~J_{\Phi}^{\mu}=\Phi^{\dagger}(\partial^{\mu}\Phi)
-(\partial^{\mu}\Phi^{\dagger}) \Phi~.$$}:
\begin{equation}
{\cal L}_{\rm Higgs}=\frac{1}{2}\partial_{\mu}H\partial^{\mu}H -V(H)~,
\label{2}
\end{equation}  
where, 
\begin{equation}
V(H)=\frac{1}{2}m_0^2 H^2 + \frac{\lambda_0}{4}H^{4}~,
\label{3}
\end{equation}
is the usual Higgs potential with tachyonic mass term, $m_0^{2}<0$. The
couplings with fermionic matter are given by the standard Lagrangian, 
\begin{equation}
{\cal L}_{\rm Higgs-Yukawa}=y^{\rm (u)}_{ij}\bar Q_L^{i} \tilde \Phi
u_R^{j} +y^{\rm (d)}_{ij}\bar Q_L^{i} \Phi d_R^{j}+
y^{\rm (l)}_{ij}\bar L_L^{i}\Phi e_R^{j}+{\rm h.c.}
\label{4}
\end{equation}
which describes the gauge-invariant interactions of the  electroweak
Higgs doublet (\ref{1}) [$\tilde \Phi=-i\tau^2\Phi^{*}$] with the
Standard Model up and down quarks ($Q^{i}_L=(u_L^{i},d^{i}_L)^{\rm T}$,
$u_R^{i}$, $d_R^{i}$; $i,j=1,2,3$ are the generation indices) and
leptons ($L_L^{i}=(\nu_L^{i},e^{i}_L)^{\rm T}$, $e_R^{i}$).  The total
Lagrangian includes also the usual gauge-invariant kinetic terms for
fermions and for $SU(3)_C$, $SU(2)_L$ and $U(1)_Y$ gauge bosons. So the
theory is indeed gauge invariant. 

There are two crucial differences between the standard theory and the
theory described by (\ref{2}):
\begin{itemize}
\item Interactions of the Higgs boson $H(x)$ with the electroweak gauge
bosons are absent in (\ref{2});
\item The "polar" degrees of freedom $\xi^{a}(x)$ are not propagating
degrees of freedom at the \emph{classical} level, since they can be
removed from the total classical Lagrangian by $SU(2)_L\times U(1)_Y$
gauge transformations.   
\end{itemize} 
Indeed, even though the Higgs field develops the (tree-level) vacuum
expectation value, 
\begin{equation}
v_0=\sqrt{-\frac{m_0^2}{\lambda_0}}\neq 0~,
\label{5}
\end{equation}
and quarks and leptons acquire their masses in the standard fashion
through (\ref{4}), the electroweak gauge bosons remain massless
classically. This means that, the would-be longitudinal degrees of
freedom $\xi^{a}(x)$ still can be rotated away. Now, since the gauge
bosons are massless (no longitudinal modes), tree-level scatterings do
not violate unitarity. However, the theory, as it stands, i.e., with
massless electroweak gauge bosons,  is obviously an incorrect theory. 

The masses for the electroweak gauge bosons emerge radiatively, through
the fermionic loops, and thus in the full quantum theory $\xi^{a}(x)$
represents propagating degrees of freedom. The mismatch of degrees of
freedom in classical and corresponding quantum theory reflects the fact
that our model is not renormalizable, and must be treated as an
effective theory. Let us see how this happens explicitly. We consider
the tree-level Lagrangian to be valid at a certain high energy scale
$\Lambda$. At lower energies $\mu < \Lambda$, the theory gets modified.
By computing the leading $\log$ one-loop contribution from the dominant
top-Higgs-Yukawa interaction, we obtain for the Higgs part of the total
Lagrangian, 
\begin{eqnarray}
\mathcal{L}_{\rm
Higgs-(1-loop)}=\frac{1}{2}\partial_{\mu}H\partial^{\mu}H
+\frac{Z}{2(1+Z)}
H^2(D_{\mu}\mathcal{X})^{\dagger}(D_{\mu}\mathcal{X})\nonumber \\
-\frac{m_H^2}{2(1+Z)}H^2- \frac{\lambda}{4(1+Z)^2}H^4~,
\label{6}
\end{eqnarray}
where $D_{\mu}=\partial_{\mu}-i\mathcal{A}_{\mu}$ is the $SU(2)\times
U(1)_Y$ covariant derivative with
$\mathcal{A}_{\mu}=gA_{\mu}^a\tau^a+g'B_{\mu}\mathbf{I}$, and
\begin{eqnarray}
Z&=&\frac{3y_t^2}{(4\pi)^2}\log\left(\Lambda^2/\mu^2\right)~,\\
m_H^2&=&m_0^2-\frac{6y_t^2}{(4\pi)^2}\left(\Lambda^2-\mu^2\right)~,\\
\lambda&=&\lambda_0+\frac{3y_t^4}{(4\pi)^2}\log\left(\Lambda^2/\mu^2\right)~.
\label{7}
\end{eqnarray}
In (\ref{6}) we have rescaled $H(x)\to H(x)/\sqrt{1+Z}$ to canonically
normalize its kinetic term\footnote{We keep the same notation for the
rescaled and original fields.}. In principle, we can renormalize the
mass and the self-interaction coupling to remove
the cut-off dependence from the Higgs potential part of 
the Lagrangian (\ref{6}). Note,
however, we are not free to renormalize  all the radiatively induced
terms in (\ref{6}). The terms not present in (\ref{2}) are
$\log$-divergent, i.e., have an explicit dependence upon $\Lambda$. This
is, of course, the manifestation of nonrenormalizability of our model. 

The effective Lagrangian (\ref{6}) is understood as an effective (Wilsonian) Lagrangian valid for the fields with momentum $|p|< \mu$, since it is obtained by integrating out fields (top-quark) with momentum $\mu <|p|<\Lambda$.\footnote{The approach here is similar to the one in \cite{Bardeen:1989ds} within the top-condensate model.} Therefore, quantum theory based on the effective Lagrangian (\ref{6}) is finite, due to the ultraviolet cut-off $\mu$. When calculating S-matrix elements using the effective Lagrangian (\ref{6}), the standard prescription is to associate $\mu$ with a total momentum of corresponding in(out)-states, i.e. $\mu=|p_{\rm in}|$. When $\mu=\Lambda$, $Z(\Lambda)=0$, and (\ref{6}) is reduced to the bare Lagrangian (\ref{2}), where the Higgs boson do not interact with the electroweak gauge bosons, and the longitudinal degrees of freedom $\xi^{a}(x)$ become non-propagating.   

Now let us use the above Lagrangian (\ref{6}) to calculate the spectrum
of the low energy theory. Fixing the unitary gauge, we obtain:
\begin{eqnarray}
v&=&\sqrt{-\frac{m_H^2}{\lambda}(1+Z)}~, \\
m_h^2 &= &\frac{-2m_H^2}{1+Z}~, \\
m_t &=& \frac{1}{\sqrt{2}}y_tv~ \text{(and similar for other
fermions)}~, \\
m_W^2&=&\frac{Z}{4(1+Z)}g^2 v^2=\frac{3g^2
m_t^2}{(32\pi^2)(1+Z)}\log\left(\Lambda^2/\mu^2\right)~,
\label{8}
\end{eqnarray}
and $m_Z^2=m_W^2\sec^2\theta_W$, where $\tan\theta_W=\frac{g'}{g}$ is
the weak mixing angle. The physical (pole) masses are defined through on-shell relations, $p^2=M^2=m^2(\mu=M)$.

We can also easily obtain interactions of the
Higgs boson with the gauge bosons:
\begin{equation}
\frac{Zg^2}{4(1+Z)}(2vh+h^2)\left[W_{\mu}^+W^{-\mu}+
\frac{Z_{\mu}Z^{\mu}}{2\cos^2\theta_W}\right]~,
\label{9}
\end{equation}
where $W_{\mu}^{\pm}=(A_{\mu}^1\mp A_{\mu}^{2})/\sqrt{2}$ and
$Z_{\mu}=\cos\theta_W A_{\mu}^3-\sin\theta_W B_{\mu}$ (in the unitary
gauge). This all looks similar to the standard theory except Z-factors
entering in the above equations. These differences give a new twist. 

Observe, that when the typical energy of the process approaches the cut-off,
$\mu \to \Lambda$, the effective mass of the gauge bosons (as well as the
interactions in (\ref{9}) go to zero, while the Higgs expectation value
(and hence the Higgs and fermion masses) remains non-zero , $v\to v_0$.
Now, if we assume that $\Lambda \sim {\rm TeV}$, the violation of
unitarity in scatterings of the longitudinal electroweak bosons can be
avoided simply because the longitudinal modes become
unphysical (non-propagating) at $\mu\equiv\sqrt{s}=\Lambda$, while the vacuum 
expectation value and the
mass of the Higgs boson can be arbitrarily large! This, of course
requires Yukawa couplings to be correspondingly small in order to fit
experimentally observed fermion masses, that is $Z$ is small. If so, the
Higgs boson can indeed be a very massive particle weakly interacting
with Standard Model fields, and thus it won't be seen at the LHC or
TEVATRON.

Unfortunately, the minimal model we have described above does give a
wrong prediction for the mass ratio:
\begin{equation}
m_W/m_t\approx 0.12~,
\label{10}
\end{equation}
(we have taken $\Lambda = 1~\mathrm{TeV}$ and
$\mu=m_t=173~\mathrm{GeV}$), while the experimental value is
$(m_W/m_t)_{\mathrm{exp}}\approx 0.46 $. 
To improve the prediction for the gauge boson masses we can introduce a
set of fermions coupled to the full electroweak Higgs doublet which give
a larger than the top-quark contribution to the gauge boson masses. 
Obviously, there are many different types of
extra hypothetical fermions which can do the job. One of the simplest
anomaly-free set is
\begin{equation}
F_L^i = (1,2,0)~,~~U_R^i =(1,1,1)~,~~D_R^i =(1,1,-1)~,
\label{11}
\end{equation}
where in paranthesis, $SU(3)$, $SU(2)$ and $U(1)_Y$
quantum numbers are indicated.  Also, we have included a generation
index $i = 1,...,N$ to allow for $N$ families of such fermions. These
fermions can have Higgs-Yukawa
interactions,
\begin{equation}
y_U^{i}\bar F_L^i \tilde \Phi U_R^i + y_D^{i}\bar F_L^i \Phi D_R^i
+\mathrm{h.c.}~,
\label{12}
\end{equation} 
where we have gone to a diagonal basis in family space.
These fermions, together with the top 
quark contribution, will radiatively generate the $W,Z$ masses
of the correct magnitude if
\begin{equation}
Z_{\rm extra}=\frac{\sum_i
y_U^{i2}+y_D^{i2}}{(4\pi)^2}\log\left(\Lambda^2/\mu^2\right) \approx 
13.5 Z_{\rm top}
\label{13}
\end{equation}
where $Z_{\rm extra}, Z_{\rm top}$ are the 1-loop contribution to the
$Z$ factor
due to the exotic fermions and top quark respectively.
This requires $\sum_i m_U^{i2} + m_D^{i2} \approx 40 m_t^2$.
Note that the exotic fermions can also have electroweak invariant
masses,
\begin{eqnarray}
{\cal L} = M_{ij} \bar F_L^i (F_L^j)^c + M'_{ij} \bar U_R^i (D_R^j)^{c} 
\end{eqnarray}
If the electroweak invariant masses are larger than the
electroweak violating masses, then  
the oblique electroweak
radiative corrections due to the exotic fermions become
suppressed \cite{Foot:1991is}, which means that  the model will
be phenomologically viable for a range of parameters.

If the Higgs mass is indeed large, this would solve the hierarchy
problem, since the radiative corrections to the ``bare'' mass would be
negligibly small. We see, that the neccessity of the low cut-off
$\Lambda \sim \mathrm{TeV}$ in our scenario is not related with a
resolution of the hierarchy problem, but is linked to an entirely
different physics. At present we are not certain what kind of theory
completes our effective description beyond the scale $\Lambda$.\footnote{One possible scenario is where 
the longitudinal degrees of freedom are composite states emergent at energies below the scale $\Lambda$, 
while the Higgs boson is elementary. The scattering of the longitudinal modes at high energies may unitarized through the corresponding form-factors. Anyway, the detailed high-energy behaviour of the scattering amplitudes 
of the longitudinal electroweak bosons would be possible to determine through the Wilsonian matching of the amplitudes of the effective theory with the corresponding amplitudes of a theory beyond the energy scale $\Lambda$. A qualitatively different mechanism of the self-unitarization of the scattering amplitudes has been recently proposed in \cite{Dvali:2010jz}, which also can be applied 
to our model \cite{arch}. } However,
our scenario can clearly be distinguished from the standard candidates
for a TeV scale physics, such as supersymmetry and technicolour, in high
energy experiments, especially through the studies of high energy
scatterings of the longitudinal modes of the massive gauge bosons. We
hope that such a new physics will be revealed in future experiments.    

Since in our scenario the vacuum expectation value is disassociated with
the 
electroweak scale, the Higgs boson can play other roles. For example,
weakly 
coupled Higgs boson can play the role of the inflaton. Also, with a
large 
vacuum expectation value it can spontaneously generate Newton's
gravitational 
constant when coupled to a scalar curvature. 
  
To conclude we have suggested a nonstandard implementation of the Higgs
boson within 
the Standard Model framework. In our scenario the Higgs boson at some energy scale
$\Lambda$ 
couples only with matter fermions, while its interactions with
electroweak gauge bosons are 
induced at lower energies radiatively. We show that the Higgs boson can
be arbitrarily 
heavy interacting with the Standard Model fields arbitrarily weakly,
while the unitarity 
in the scatterings of the longitudinal electroweak gauge bosons is
maintaned if $\Lambda \sim \mathrm{TeV}$.

\paragraph{Acknowledgments.} 
 This work was supported by the Australian Research Council.



\end{document}